\documentclass[12pt,a4paper]{article}
\usepackage{a4wide}
\usepackage{amsmath}
\usepackage{amssymb}
\usepackage{epsfig}
\usepackage{subfigure}
\usepackage{exscale}
\usepackage{float}
\usepackage{bbm}
\usepackage[numbers,sort&compress]{natbib}
\usepackage{pst-plot, pstricks,pst-math}
\usepackage{fancybox,amssymb,color}
\usepackage{graphicx}
\usepackage{pstricks, color, graphicx, epsfig, psfrag}
\usepackage{amsfonts,amsmath,amssymb,slashed}
\usepackage{dsfont}
\usepackage{bbm,bm}
\usepackage{fancyhdr, a4wide}
\usepackage[english]{babel}
\usepackage{subfigure}

\makeatletter
\@addtoreset{equation}{section}
\makeatother

\title{Goldstone Boson Interaction in D=2+1 (Pseudo-)Lorentz-Invariant Systems with a Spontaneously Broken Internal Rotation Symmetry}

\author{Christoph P.\ Hofmann$^a$ \\ \\
\normalsize {$^a$ Facultad de Ciencias, Universidad de Colima} \\
\vspace{0.3cm}
\normalsize {Bernal D\'iaz del Castillo 340, Colima C.P.\ 28045, Mexico} \\}

\begin{document}

\maketitle

\begin{abstract} \normalsize

The low-temperature properties of systems characterized by a spontaneously broken internal rotation symmetry, O($N$) $\to$ O($N$-1), are
governed by Goldstone bosons and can be derived systematically within effective Lagrangian field theory. In the present study we consider
systems living in two spatial dimensions, and evaluate their partition function at low temperatures up to three-loop order. Although our
results are valid for any such system, here we use magnetic terminology, i.e., we refer to quantum spin systems. We discuss the sign of
the Goldstone boson interaction in the pressure, staggered magnetization, and susceptibility as a function of an external staggered field
for general $N$. As it turns out, the $d$=2+1 quantum XY model ($N$=2) and the $d$=2+1 Heisenberg antiferromagnet ($N$=3), are rather
special, as they represent the only cases where the spin-wave interaction in the pressure is repulsive in the whole parameter regime where
the effective expansion applies. Remarkably, the $d$=2+1 XY model is the only system where the interaction contribution in the staggered
magnetization (susceptibility) tends to positive (negative) values at low temperatures and weak external field.

\end{abstract}

\newpage


\maketitle

\section{Introduction}
\label{Intro}

The present study is devoted to (pseudo-)Lorentz-invariant systems that are defined in two spatial dimensions and -- at $T$=0 -- are
characterized by a spontaneously broken rotation symmetry O($N$) $\to$ O($N$-1). We focus on the low-energy dynamics which is dominated by
the corresponding $N$-1 Goldstone bosons. The most prominent physical realizations are the $d$=2+1 quantum XY model ($N$=2) and the
$d$=2+1 Heisenberg antiferromagnet ($N$=3). In either case the relevant low-energy degrees of freedom are the spin waves or magnons.

Although there exists a vast number of publications -- in particular for $N$=\{2,3\} -- the impact of the Goldstone boson interaction onto
the low-temperature behavior of these systems has been largely neglected. While the specific cases $N$=\{2,3\} have been addressed within
effective Lagrangian field theory in Refs.~\citep{CHN89,NZ89,Fis89,HL90,HN91,HN93,Hof10,Hof14a}, a systematic study of the nature of the
interaction in systems with general $N$, seems to be lacking. It is the aim of the present investigation to rigorously analyze the effect
of the Goldstone boson interaction in the pressure, in the staggered magnetization, and in the susceptibility, using the systematic and
model-independent effective Lagrangian technique.

A method widely used to analyze the low-temperature behavior of spin systems with a spontaneously broken internal rotation symmetry is
spin-wave theory, or its modified versions adapted to two spatial dimensions \citep{UTT79,Tak86,GJ87,Tak87a,Tak89a,Tak89b,OY89,OY90,Tak92,
KSK03}. With respect to the $d$=2+1 quantum XY model, the interested reader may also consult Refs.~\citep{HOW91,WOH91,Zha93,HHB99}. The
finite-temperature properties of the $d$=2+1 Heisenberg antiferromagnet and the $d$=2+1 quantum XY model, have furthermore been analyzed
with alternative methods that include Schwinger-boson mean-field theory, Green's functions, renormalization group methods, and Monte-Carlo
simulations. Concerning the $d$=2+1 quantum XY model we refer to
Refs.~\citep{RR84,Din92,Mak92,JY94,Pir96,SH99,TR01,SFBI01,MSS04,CNPV13,RD13}, while for the $d$=2+1 Heisenberg antiferromagnet we point to
Refs.~\citep{CHN88,AA88a,AA88b,GJN89,Bar91,Man91,CSY94,WY94,San97,CTVV97,BBGW98,KT98,MS00,SS01}. For the general case, i.e., for arbitrary
$N$, see Refs.~\citep{HK79,FP85,Has90,GL91a,GL91b,GL91c,DNJN91,IK97,BNPWW12}.

In the work reported here, we do not make use of spin-wave theory or any other microscopic method. Rather, our approach relies on the
effective Lagrangian technique which systematically captures the low-energy physics of the system that is governed by the Goldstone
bosons. Low energy -- or low temperature -- means that we are interested in the properties of the system below the intrinsic scale defined
by the underlying microscopic model. In connection with the Heisenberg antiferromagnet or the quantum XY model, this scale is given by the
exchange coupling $J$. Note that the spontaneously broken rotation symmetry O($N$) is only approximate, since we incorporate a weak
external field into the low-energy description. As we discuss below, this field cannot be completely switched off in two spatial
dimensions in the effective approach we pursue in the present study.

The $d$=2+1 quantum XY model ($N$=2) and the $d$=2+1 Heisenberg antiferromagnet ($N$=3) turn out to be rather peculiar cases: they
represent the only systems where the spin-wave-interaction contribution in the pressure is repulsive at low temperatures in the entire
parameter regime where our effective analysis applies. Furthermore, considering the impact of the spin-wave interaction in the staggered
magnetization and susceptibility, the low-temperature behavior of the $d$=2+1 quantum XY model is quite different from any other $d$=2+1
(pseudo-)Lorentz-invariant system with a spontaneously broken internal rotation symmetry O($N$): it is the only system where the
interaction contribution in the staggered magnetization (susceptibility) tends to positive (negative) values at low temperatures and weak
external field. Interestingly, for systems with $N \ge 4$ the staggered magnetization (susceptibility) also becomes positive (negative) at
very low temperatures and not too weak external fields. This means that here the Goldstone boson interaction at finite temperature tends
to reinforce the antialignment of the spins and enhances the staggered spin pattern.

We emphasize that the physical realizations for $N$=\{2,3\} are not restricted to quantum magnetism. Rather, any two-dimensional system
with a spontaneously broken rotation symmetry, O(3) $\to$ O(2), or O(2) $\to$ 1, exhibits the same characteristic features determined by
the universal physics of its Goldstone bosons. Most remarkably, the question of whether the Goldstone boson interaction in the pressure
and the other observables we consider, is repulsive or attractive at low temperatures, applies to any $d$=2+1 (pseudo-)Lorentz-invariant
system that exhibits a spontaneously broken rotation symmetry. In this sense, our results on the nature of the Goldstone boson interaction
are universal. To the best of our knowledge, our three-loop results are far beyond any analogous calculation based on more conventional
condensed matter methods like spin-wave theory. 

Regarding the $d$=2+1 quantum XY model and the Heisenberg antiferromagnet, high-precision measurements based on very efficient loop-cluster
algorithms have been performed recently in Refs.~\citep{GHJNW09,GHJPSW11}. These numerical results were confronted with analytic
calculations obtained within effective Lagrangian theory. Not only were some combinations of low-energy constants determined at the
permille level, but also was the correctness of the systematic effective Lagrangian method demonstrated.

In order to convince the interested reader that the effective Lagrangian technique indeed represents a very powerful machinery in the
condensed matter domain, we point out the following additional references which are all based on that method. Ferromagnets and
antiferromagnets in one, two, and three spatial dimensions have been analyzed in
Refs.~\citep{Leu94a,Hof99a,Hof99b,Hof01,Hof02,Hof11a,RS99a,RS99b,RS00,GHKW10,Hof11b,Hof11c,Hof12a,Hof12b,Hof13a,RPPK13,Hof14b,Hof14c,
ABHV14,Rad15}. (Quasi-)two-dimensional antiferromagnets that turn into high-temperature superconductors upon doping with either holes or
electrons, have been systematically studied in
Refs.~\citep{KMW05,BKMPW06,BKPW06,BHKPW07,BHKMPW07,BHKPW08,BHKMPW09,JKHW09,KBWHJW12,VHJW12,JKBWHW12,VHJW15}.

The rest of the paper is organized as follows. In Sec.~\ref{EffectiveTheory} we give a concise account of the effective Lagrangian method.
The three-loop analysis of the partition function is reviewed in Sec.~\ref{FreeEnergyDensity}. The low-temperature series for the
pressure, the staggered magnetization, and the susceptibility for $d$=2+1 (pseudo-)Lorentz-invariant systems with a spontaneously broken
symmetry, O($N$) $\to$ O($N$-1), are discussed in Sec.~\ref{GeneralN}. Our focus is the manifestation of the Goldstone boson interaction
in the low-temperature behavior of these systems in a weak external field. Finally, in Sec.~\ref{Summary} we present our conclusions.

\section{Effective Field Theory}
\label{EffectiveTheory}

The basic principles underlying the effective Lagrangian method, as well as the perturbative evaluation of the partition function, have
been outlined on various occasions and will not be repeated here in detail. Rather, in this section, we discuss some essential aspects of
the method, such that a non-expert can understand the calculation presented here. The interested reader may find details in Appendix A of
Ref.~\citep{Hof11a} or in Section 2 of Ref.~\citep{Hof12a}. Moreover, for pedagogic introductions to the effective Lagrangian technique we
refer to Refs.~\citep{Brau10,Bur07,Goi04,Sch03,Leu95,Eck95}.

The basic degrees of freedom in the effective Lagrangian are the $N$-1 Goldstone bosons, which result from the spontaneously broken
rotation symmetry O($N$) $\to$ O($N$-1). These Goldstone boson fields are collected in a $N$-dimensional unit vector $U^i(x)$
\begin{equation}
U^i(x) \, U^i(x) \, = \, 1 \, , \qquad i=1,2, \dots, N \, .
\end{equation}
While the first $N$-1 components describe the Goldstone bosons, the last component,
\begin{equation}
U^N(x) = \sqrt{1-U^a(x) U^a(x)} \, , \qquad a=1,2, \dots, N-1 \, ,
\end{equation}
guarantees that the vector ${\vec U} = U^i(x)$ is normalized to unity.

The low-energy dynamics of the system, dominated by the Goldstone bosons, is described by the following Euclidean effective Lagrangian:
\begin{eqnarray}
\label{Leff}
{\cal L}_{eff} & = & {\cal L}^2_{eff} + {\cal L}^4_{eff} \, , \nonumber \\
{\cal L}^2_{eff} & = & \mbox{$ \frac{1}{2}$} F^2 {\partial}_{\mu}
U^i{\partial}_{\mu} U^i - {\Sigma}_s H^i_s U^i \, , \nonumber\\
{\cal L}^4_{eff} & = & - e_1 ({\partial}_{\mu} U^i
{\partial}_{\mu} U^i)^2 - e_2 \, ({\partial}_{\mu} U^i {\partial}_{\nu}
U^i)^2 + k_1 \frac{{\Sigma}_s}{F^2} \, (H^i_s U^i)
({\partial}_{\mu} U^k {\partial}_{\mu} U^k) \nonumber \\
& & - k_2 \frac{{\Sigma}_s^2}{F^4} \, (H^i_s U^i)^2 - k_3 \frac{{\Sigma}_s^2}{F^4} \, H^i_s H^i_s \, .
\end{eqnarray}
The different pieces in ${\cal L}_{eff}$ are organized according to the number of derivatives they contain. The leading contribution
${\cal L}^2_{eff}$ contains two space-time derivatives (order $p^2$), while the next-to-leading contribution ${\cal L}^4_{eff}$ is of order
$p^4$. The external field $H^i_s$ counts as a quantity of order $p^2$.

A physical realization for $N$=2 and $N$=3 is provided by the $d$=2+1 quantum XY model and the $d$=2+1 Heisenberg antiferromagnet,
respectively. The O(2) spin symmetry of the quantum XY Hamiltonian, and the O(3) spin symmetry of the Heisenberg Hamiltonian, are
spontaneously broken according to O(2) $\to$ 1 and O(3) $\to$ O(2), respectively. The Goldstone bosons in these cases are identified with
the spin waves or magnons, and the external field ${\vec H_s} = (0,\dots,0,H_s)$ is the staggered field that couples to the staggered
magnetization vector $\vec U$. Note that there exists a mapping between the antiferromagnetic and the ferromagnetic quantum XY model on a
bipartite lattice, such that for $N$=2 the external field may also be interpreted as the magnetic field that couples to the magnetization
vector $\vec U$ of the quantum XY ferromagnet.

It should be stressed that the effective description is universal, in the sense that any two (pseudo-)Lorentz-invariant systems that
display the same internal and space-time symmetries, will be described by the same effective Lagrangian. But while the structure of the
effective Lagrangian is the same, the difference between two distinct physical realizations manifests itself in the actual numerical
values of the low-energy coupling constants $F,{\Sigma}_s,e_1, e_2, k_1, k_2, k_3$. Symmetry leaves these effective constants
undetermined, such that they have to be extracted from a numerical simulation, by comparison with the microscopic calculation, or by
experiment. In the context of quantum spin systems, $F^2$ is referred to as helicity modulus or spin stiffness, while ${\Sigma}_s$ is the
staggered magnetization at zero temperature.

An important point to emphasize is that the effective Lagrangian, Eq.~(\ref{Leff}), is written in a Lorentz-invariant form. One may thus
be very skeptical about its applicability to the $d$=2+1 quantum XY model or the Heisenberg antiferromagnet, whose underlying
two-dimensional lattices are not even invariant under continuous space rotations. First of all, the leading piece ${\cal L}^2_{eff}$ does
not depend on the specific geometry -- the lattice structure only starts manifesting itself at subleading orders in the effective
Lagrangian \cite{HN91}. Hence ${\cal L}^2_{eff}$ may be written in a (pseudo)-Lorentz-invariant form, where the spin-wave velocity takes
over the role of the velocity of light. Note that (pseudo-)Lorentz-invariance is an accidental symmetry of ${\cal L}^2_{eff}$, and is not
present in the microscopic XY or Heisenberg Hamiltonian. As far as the next-to-leading piece ${\cal L}^4_{eff}$ is concerned, indeed,
lattice anisotropies start to show up at this order: these contributions are still invariant under the discrete spatial symmetries of the
lattice, but break continuous space rotation invariance and thus (pseudo-)Lorentz invariance. As will become clear in the next section,
these extra contributions do not affect our conclusions concerning the impact of the spin-wave interaction in thermodynamic quantities. In
other words, it is perfectly safe for our purposes to also write ${\cal L}^4_{eff}$ in a (pseudo-)Lorentz-invariant form.

Since we are dealing with relativistic kinematics, the external field $H_s$ induces a mass term -- or energy gap -- in the dispersion
relation of the Goldstone bosons,
\begin{equation}
\label{LeadingMassTerm}
\omega = \sqrt{v^2 {\vec k}^2 + v^4 M^2} \, , \qquad M^2 = \frac{{\Sigma}_s H_s}{F^2} \, ,
\end{equation}
where $v$ is the spin-wave velocity. The low-energy degrees of freedom are then referred to as pseudo-Goldstone bosons. As long as the
external field $H_s$, and therefore $M^2$, are small compared to the intrinsic scale occurring in the underlying microscopic model, these
quantities may be treated as perturbations within the effective Lagrangian framework. In the case of the quantum XY model or the
Heisenberg antiferromagnet, this scale is given by the exchange integral $J$ that appears in the corresponding microscopic Hamiltonians.

\section{Three-Loop Analysis of the Partition Function}
\label{FreeEnergyDensity}

The evaluation of the partition function for (pseudo-)Lorentz-invariant systems, defined in two spatial dimensions and exhibiting a
spontaneously broken symmetry O($N$) $\to$ O($N$-1), has been presented in Ref.~\citep{Hof10} within effective Lagrangian field theory up
to the three-loop level. In order not to be repetitive, here we restrict ourselves to a minimum of information -- just to understand the
subsequent discussion.

\begin{figure}
\includegraphics[width=15cm]{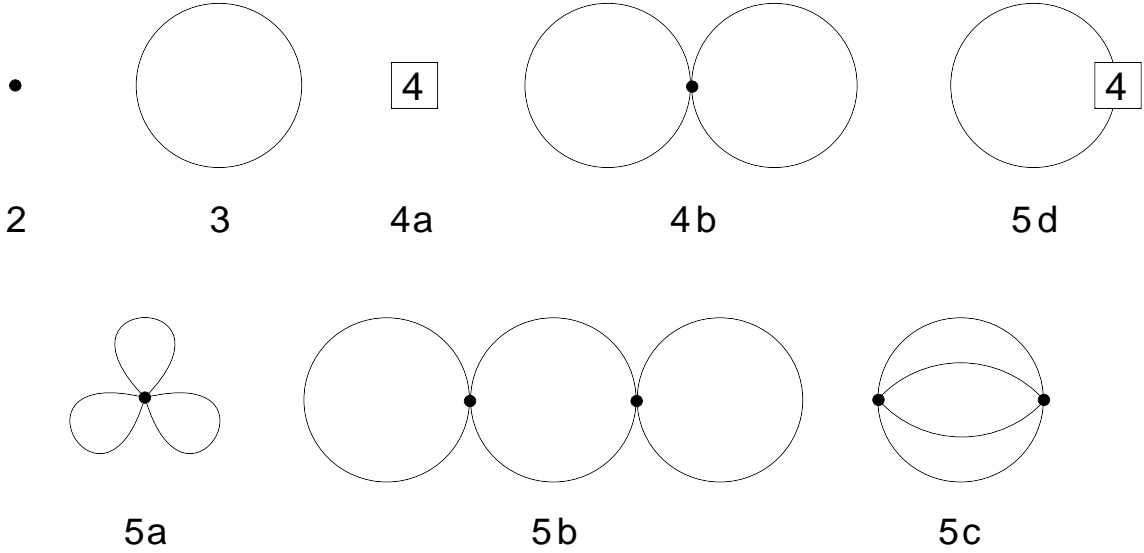}
\caption{Feynman diagrams concerning the low-temperature expansion of the partition function for systems with a spontaneously broken
rotation symmetry O($N$) $\to$ O($N$-1), up to three-loop order and for two spatial dimensions. The number 4 attached to the vertices in
graphs 4a and 5d, indicates that they originate from ${\cal L}^4_{eff}$. Vertices involving the leading piece ${\cal L}^2_{eff}$ are
denoted by a filled circle. In $d$=2+1, each Goldstone boson loop is suppressed by one momentum (or temperature) power.}
\label{figure1}
\end{figure}

Up to three-loop order, one has to evaluate the eight Feynman diagrams depicted in Fig.\ref{figure1}. We should stress that we are dealing
with a systematic loop expansion of the partition function, where each Goldstone boson loop is suppressed by one power of momentum. The
two-loop graph 4b, compared to the one-loop graph 3, is suppressed by one additional power of $p$. The three-loop graphs 5a-c, on the
other hand, are suppressed by two additional powers of momentum compared to graph 3, such that they are of next-to-next-to leading order
$p^5$. While the evaluation of most of these graphs turns out to be straightforward within effective field theory, the piece of resistance
is the three-loop graph 5c. The renormalization and numerical evaluation of this graph has been presented in detail in
Refs.~\citep{Hof10,Hof14a}.

The final result for the free energy density ($N \ge 2$) up to three-loop order is
\begin{eqnarray}
\label{fedTau}
z & = & z_0 - \mbox{$\frac{1}{2}$} (N-1) h_0(\sigma) \, T^3
\, + \, \mbox{$\frac{1}{8}$} (N-1) (N-3) \frac{1}{F^2 {\tau}^2} {h_1(\sigma)}^2
\, T^4 \nonumber \\
& - & \mbox{$\frac{1}{128 \pi}$} (N-1) (N-3) (5N-11) \frac{1}{F^4 {\tau}^3}
{h_1(\sigma)}^2 \, T^5 \nonumber \\
& + & \mbox{$\frac{1}{48}$} (N-1) (N-3) (3N-7) \frac{1}{F^4 {\tau}^2}
{h_1(\sigma)}^3 \, T^5 \nonumber \\
& - & \mbox{$\frac{1}{16}$} (N-1) {(N-3)}^2 \frac{1}{F^4 {\tau}^4}
{h_1(\sigma)}^2 h_2(\sigma) \, T^5
\, + \, \frac{1}{F^4} \, q(\sigma) \, T^5 + {\cal O}(T^6) \, ,
\end{eqnarray}
where the ratios
\begin{equation}
\tau = \frac{T}{M_{\pi}} \, , \qquad \sigma = \frac{M_{\pi}}{2 \pi T} \, ,
\end{equation}
define dimensionless variables, and $z_0$ is the free energy density at zero temperature. The quantity $M_{\pi}$ is the renormalized
mass of the $N$-1 pseudo-Goldstone bosons,
\begin{eqnarray}
\label{MassRen}
M^2_{\pi}(H_s) & = & \frac{{\Sigma}_s H_s}{F^2} - \frac{(N-3)}{8 \pi} \frac{{\Sigma}^{3/2}_s H_s^{3/2}}{F^5} \nonumber \\
& + & \Big\{ 2 (k_2 - k_1) + \frac{b_1}{F^2} + \frac{b_2}{64 \pi^2 F^2}
\Big\} \frac{{\Sigma}^2_s H_s^2}{F^6} \nonumber \\
& + & {\cal O}(H_s^{5/2}) ,
\end{eqnarray}
or, equivalently, the energy gap in the presence of the external field $H_s$. Note that the first term on the RHS is the leading mass
contribution (order $p^2 \propto H_s$) defined in Eq.~(\ref{LeadingMassTerm}), while the remaining two contributions in 
Eq.~(\ref{MassRen}) are subleading corrections of order $p^3$ and $p^4$, respectively. The quantities $k_1$ and $k_2$ are the
next-to-leading-order effective constants defined in Eq.~(\ref{Leff}), and the explicit expressions for the coefficients $b_1$ and $b_2$
are listed in Appendix B of Ref.~\citep{Hof10}.

\begin{figure}
\begin{center}
\includegraphics[width=14cm]{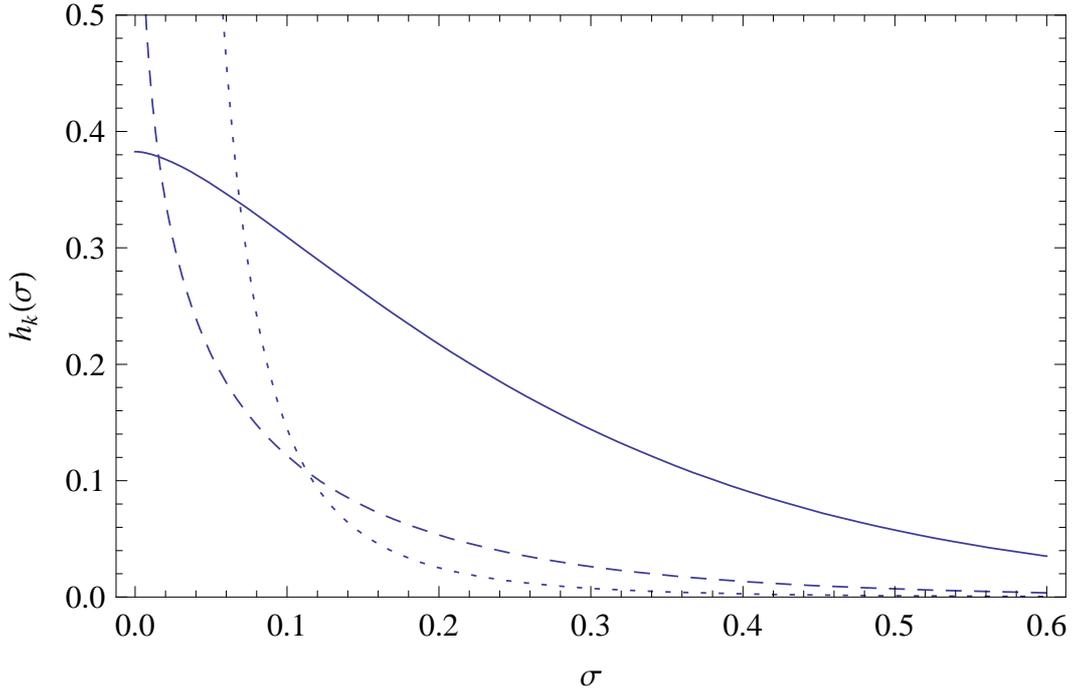}
\end{center}
\caption{The kinematical functions $h_k(\sigma)$: $k$=0 (continuous curve), $k$=1 (dashed curve), and $k$=2 (dotted curve), in terms of
the parameter $\sigma = M_{\pi}/2\pi T$.}
\label{kinematical}
\end{figure}

The dimensionless kinematical functions $h_0, h_1$ and $h_2$ appearing in Eq.~(\ref{fedTau}), are proportional to Bose functions
$g_r(M,T)$ in $d$ space-time dimensions,
\begin{equation}
\label{ThermalDimensionless}
g_0 = T^3 \, h_0(\sigma) \, , \qquad g_1 = T \, h_1(\sigma) \, , \qquad g_2 = \frac{1}{T} \, h_2(\sigma) \, ,
\end{equation}
which are defined as
\begin{equation}
\label{BoseFunctions}
g_r(M,T) \, = \, 2 {\int}_{\!\!\! 0}^{\infty}
\frac{\mbox{d} \rho}{{(4 \pi \rho)}^{d/2}} \, {\rho}^{r-1}
\, \exp(- \rho M^2) \, \sum_{n=1}^{\infty} \, \exp(-n^2 / 4 \rho T^2) \, ,
\end{equation}
and which we have depicted in Fig.~\ref{kinematical}. Note that in the limit $\sigma \to 0$, the functions $h_1$ and $h_2$ diverge,
whereas $h_0$ tends to a finite value.

Finally, the function $q$ appearing in the last term of Eq.~(\ref{fedTau}) is given by
\begin{eqnarray}
\label{definitionI}
T^5 q(\sigma) & = & \mbox{$ \frac{1}{48}$} (N-1) (N-3) \, M_{\pi}^4 \, {\bar J}_1 - \mbox{$ \frac{1}{4}$} (N-1) (N-2) \, {\bar J}_2 \, ,
\nonumber \\
& = & \mbox{$ \frac{1}{48}$} (N-1) (N-3) \, q_1(\sigma) \, T^5 - \mbox{$ \frac{1}{4}$} (N-1) (N-2) \, q_2(\sigma)  \, T^5 \, .
\end{eqnarray}
\begin{figure}
\begin{center}
\includegraphics[width=12cm]{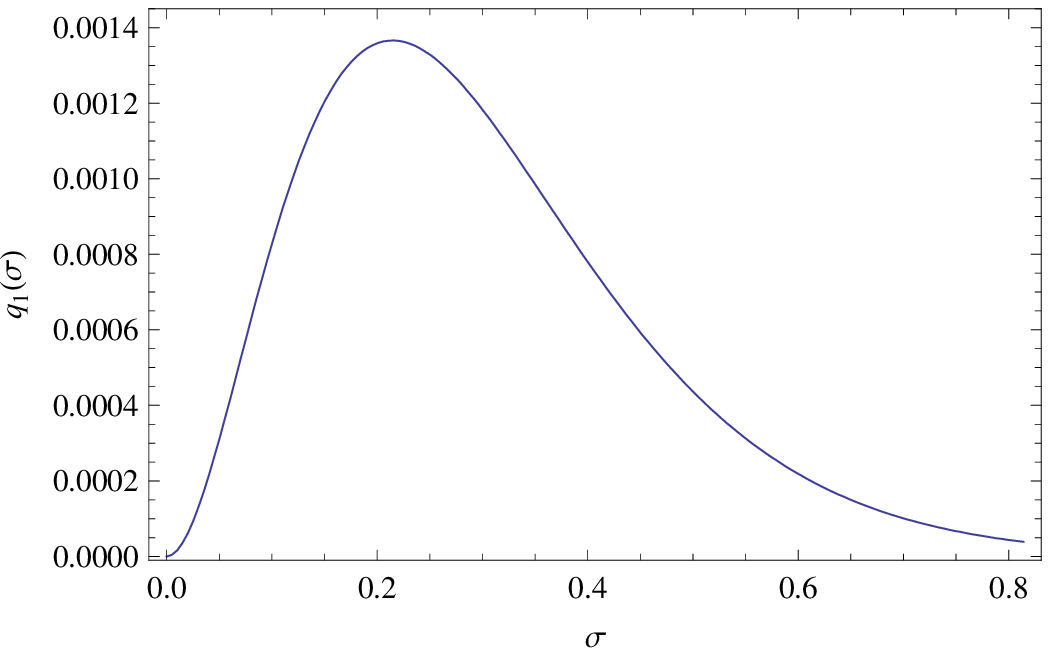}

\vspace{5mm}

\includegraphics[width=12cm]{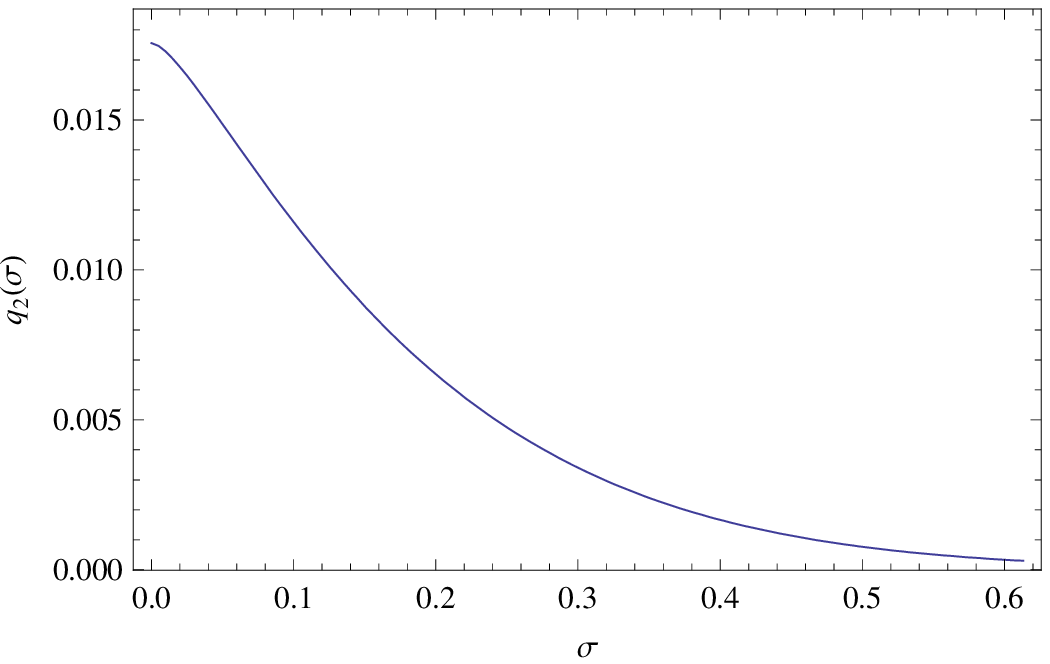}
\end{center}
\caption{The three-loop functions $q_1(\sigma)$ (top) and $q_2(\sigma)$ (bottom), in terms of the parameter $\sigma = M_{\pi}/2\pi T$.}
\label{functionsq1q2}
\end{figure}
A plot for the two individual functions $q_1$ and $q_2$, in terms of the parameter $\sigma=M_{\pi}/2\pi T$, is provided in
Fig.~\ref{functionsq1q2}. The function $q$ involves the renormalized integrals ${\bar J}_1$ and ${\bar J}_2$,
\begin{eqnarray}
\label{decompositionJ}
{\bar J}_1 & = & J_1 - c_1 - c_2 \, g_1(M,T) \, , \nonumber \\
{\bar J}_2 & = & J_2 - c_3 - c_4 \, g_1(M,T) \, , 
\end{eqnarray}
stemming from the three-loop graph 5c. These finite integrals are related to the unrenormalized integrals $J_1$ and $J_2$ over the torus
${\cal T} = {\cal R}^{d-1} \times S^1$ via
\begin{eqnarray}
J_1 & = & {\int}_{\!\!\! {\cal T}} {\mbox{d}}^d x \, \Big\{ G(x) \Big\}^4 \, ,
\nonumber \\
J_2 & = & {\int}_{\!\!\! {\cal T}} {\mbox{d}}^d x \,
\Big\{ \partial_{\mu} G(x) \, \partial_{\mu} G(x) \Big\}^2 \, .
\end{eqnarray}
The coefficients $c_1, \dots, c_4$ represent subtraction constants which remove the infinities in the integrals $J_1$ and $J_2$ -- for the
explicit expressions the reader may consult Refs.~\citep{Hof10,Hof14a}.

The low-temperature series for the free energy density of (pseudo-)Lorentz-invariant systems, defined in two spatial dimensions and
displaying a spontaneously broken symmetry O($N$) $\to$ O($N$-1), thus exhibits the following general structure. The leading contribution
is of order $T^3$ -- this is the free Bose gas term which originates from the one-loop graph 3. The interaction sets in at order $T^4$ and
corresponds to the two-loop graph 4b. We then have four interaction contributions of order $T^5$ that come from graphs 5a-c.

Note that the one-loop graph 5d that describes noninteracting Goldstone bosons, is the only loop graph that involves an insertion from
${\cal L}^4_{eff}$ where lattice anisotropies start to show up -- it is important to stress that the low-temperature structure of the
interaction terms is therefore not affected by these anisotropies. There exists, however, a subtle effect as the one-loop graph 5d
manifests itself in the renormalized Goldstone boson mass $M_{\pi}$ given in Eq.~(\ref{MassRen}): the combination $k_2$ - $k_1$ of
next-to-leading effective constants in the curly bracket originates from graph 5d, indicating that lattice anisotropies start showing up
at this order. But note that we are dealing with a next-to-next-to-leading order effect in a temperature-{\it independent} quantity which
does not affect our conclusions regarding the nature of the Goldstone boson interaction.

\section{Nature of Goldstone Boson Interaction}
\label{GeneralN}

Let us explain what we mean by {\it nature} of the Goldstone boson interaction. First of all, we refer to a diagrammatic definition of the
interaction. Tree graphs or one-loop graphs related to the partition function describe free Goldstone bosons. But any partition function
diagram that contains two or more loops, represents an interaction diagram: here, four or more Goldstone boson lines meet at one vertex.
In the present context of $d$=2+1 (pseudo-)Lorentz-invariant systems with a spontaneously broken rotation symmetry, the interaction
manifests itself at two-loop order $p^4$ (diagram 4b) and at three loop-order $p^5$ (diagrams 5a-c).

Then we adopt a thermodynamically motivated interpretation of the interaction, which refers to the sign of the interaction contribution in
the low-temperature series of the various quantities we consider -- this is what we mean by {\it nature} of Goldstone boson interaction.
Regarding the pressure, a positive (negative) interaction contribution is interpreted as repulsive (attractive), since it is the picture
of the non-ideal Goldstone boson gas we have in mind. Concerning the low-temperature series for the staggered magnetization (order
parameter) and the staggered susceptibility, the sign of the interaction contribution is related to the antialignment of the spins, as we
will discuss.

\subsection{Effective Expansion: Domain of Validity}
\label{ParameterRange}

Before submerging into the analysis of the interaction, we should point out in which parameter regime -- defined by $T$, $H_s$ (or
$M_{\pi}$), $F$, $J$ -- the effective series are valid. To explore this domain, we introduce the two dimensionless ratios
\begin{equation}
\label{domRatios}
\frac{T}{2 \pi F^2} \, , \qquad \frac{H_s}{2 \pi F^2} \, .
\end{equation}
Note that, at leading order, the external field $H_s$ is proportional to the square of the Goldstone boson mass $M_{\pi}$ according to
Eq.~(\ref{MassRen}).

For the effective expansion to be valid, both the temperature and the external field must be small with respect to the underlying scale of
the system. In chiral perturbation theory, i.e., the effective field theory of quantum chromodynamics (QCD), this underlying scale --
the chiral symmetry breaking scale ${\Lambda}_{QCD}$ -- is of the order of ${\Lambda}_{QCD} = 4 \pi {\cal F}$, where
${\cal F} \approx 93 MeV$ is the pion decay constant. In the present study, we adopt this definition of the scale, bearing in mind that in
connection with spin systems with a spontaneously broken rotation symmetry (and in two spatial dimensions), the analog of the pion decay 
constant is the helicity modulus or spin stiffness $F^2$. This quantity has been measured in high-precision Monte Carlo simulations, both
for the $d$=2+1 quantum XY model and the $d$=2+1 Heisenberg antiferromagnet on the square lattice, with the result
\citep{JY94,BBGW98,GHJPSW11}:
\begin{equation}
F^2_{XY} = 0.26974(5) J \, , \quad F^2_{AF} = 0.1808(4) J \, .
\end{equation}
Here the underlying scale -- the exchange constant $J$ of the microscopic theory -- can thus be expressed in terms of the effective
coupling constant $F$ as $J \approx 2 \pi F^2$ \citep{Hof10}. Accordingly, the two dimensionless quantities defined in
Eq.~(\ref{domRatios}) simply are
\begin{equation}
\frac{T}{J} \, , \qquad \frac{H_s}{J} \, .
\end{equation}

In order to determine what "low temperature" means, let us again consider the situation in quantum chromodynamics. Unlike in the present
case where $d$=2+1, in QCD we are dealing with three spatial dimensions. QCD is characterized by a chiral phase transition taking place at
$T_{QCD} \approx 190 MeV$ where the Goldstone boson picture definitely ceases to be valid. Compared to the scale of QCD,
$\Lambda_{QCD} \approx 1 GeV$, we obtain the ratio
\begin{equation}
r_{\mbox{\tiny T}} = \frac{T_{QCD}}{\Lambda_{QCD}} \approx 0.2 \, .
\end{equation}
If we define low temperature in relation to the critical temperature, let us say, $T \lesssim 0.4 \, T_{QCD}$, in terms of the QCD scale, 
low temperature means
\begin{equation}
T \lesssim 0.08 \, \Lambda_{QCD} \approx 0.08 \cdot 4 \pi {\cal F} \, .
\end{equation}
Returning to our systems in $d$=2+1 with a spontaneously broken rotation symmetry, we conclude that the low-temperature region where the
effective expansion applies, is given by 
\begin{equation}
\label{rangeTJ}
\boxed{T \lesssim 0.08 \, J \approx 0.08 \cdot 2 \pi F^2 \, .}
\end{equation}
In order to check consistency, consider the $d$=2+1 quantum XY model that is characterized by the Kosterlitz-Thouless phase transition
which marks the breakdown of the spin-wave picture. On the square lattice, the Kosterlitz-Thouless transition occurs at $T_{KT}=0.343 J$
\citep{HK98}. If we define low temperature again in relation to the critical temperature, $T \lesssim 0.4 \, T_{KT}$, in terms of the
microscopic scale $J$, we get $T \lesssim 0.1 \, J$ which is perfectly consistent with Eq.~(\ref{rangeTJ}).

Regarding the magnitude of the Goldstone boson mass (and hence the external field $H_s$), recall that in QCD the pion masses are roughly
$M_{\Pi} \approx 140 MeV$, leading to the ratio
\begin{equation}
r_{\mbox{\tiny M}} = \frac{M_{\Pi}}{\Lambda_{QCD}} \approx 0.1 \, .
\end{equation}
In fact, the effective theory of QCD even works for kaon masses ($M_K \approx 500 MeV$) and the $\eta$-mass ($M_{\eta} \approx 550 MeV$)
-- these particles, much like the pions, represent Goldstone bosons. We thus conclude that the effective series derived in the present
paper are certainly valid in the region
\begin{equation}
\boxed{H_s, M_{\pi} \, \lesssim 0.1 \, J \approx 0.1 \cdot 2 \pi F^2 \, ,}
\end{equation}
which defines weak external field $H_s$, or low mass $M_{\pi}$. Again, the two quantities are connected by Eq.~(\ref{MassRen}). Note that
the Goldstone boson mass is given in units of $[ M_{\pi} v^2 ] = J$, where the spin-wave velocity $v$ has been set to one.

However, there is an important caveat: although the ratio $H_s/J$ must be small, it cannot be arbitrarily small, because of the subtleties
arising in two spatial dimensions.
\begin{figure}
\includegraphics[width=14cm]{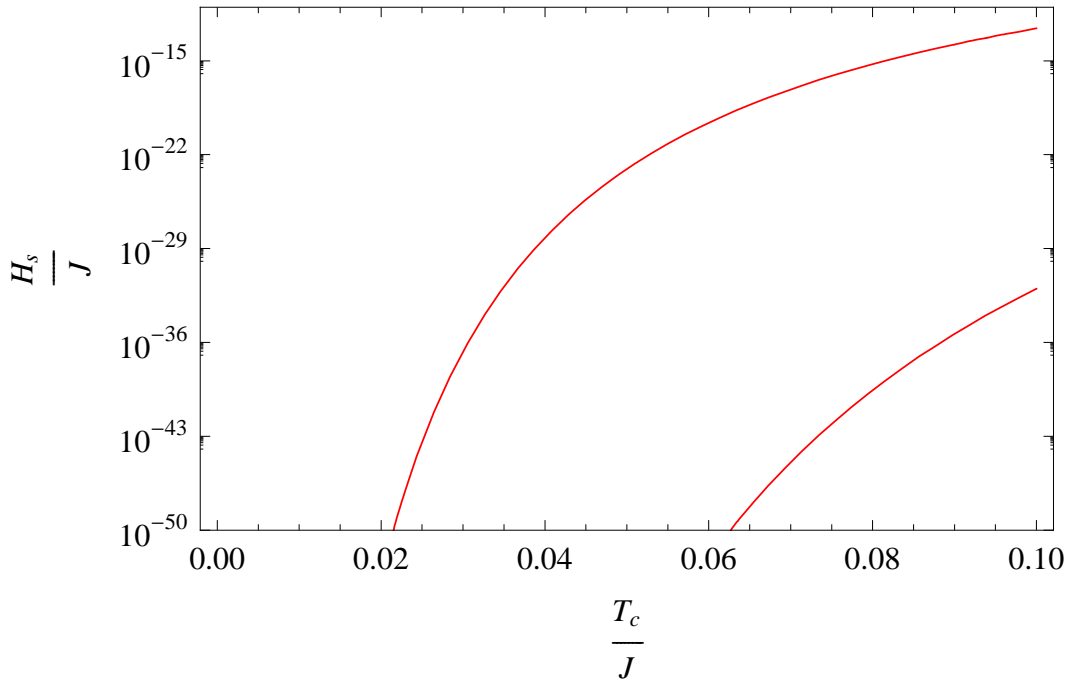}
\caption{Domain of validity of the effective low-temperature expansion of the square lattice quantum XY model (bottom) and the square
lattice Heisenberg antiferromagnet (top). The effective expansion is safe in the parameter region above these curves.}
\label{figure2}
\end{figure}
In the absence of an external field, spontaneous symmetry breaking does not take place in two spatial dimensions at finite temperature
(Mermin-Wagner theorem \citep{MW66}). Formally this means that the limit $H_s \to 0$ in the low-temperature series is problematic, as we
have discussed on various occasions \citep{Hof10,Hof14a}: if the external field is very weak, we leave the domain of validity of the
present effective expansion. Still, the excluded region in the plane $H_s/J$ versus $T/J$ is tiny, as we illustrate in Fig.~\ref{figure2}
for the square lattice quantum XY model and the Heisenberg antiferromagnet. In the parameter region above these curves, the effective
analysis presented here perfectly applies. In Sec.~\ref{InteractionOrderParameter} where we consider the staggered magnetization, we
explain how we have generated these curves.

In order to illustrate the restrictions imposed by very small Goldstone boson masses for general systems (not necessarily spin models
where the $M_{\pi}$ is determined by the staggered field), we follow another strategy. According to Ref.~\citep{HN90,DHJS91}, for
$N \ge 3$ a nonperturbative mass gap $m$ is generated for the Goldstone bosons at finite temperature for $d$=2+1 dimensional O($N$)
symmetric theories,
\begin{equation}
\label{massGap}
m = {\Big( \frac{8}{e} \Big)}^X \, \frac{T}{\Gamma(1+X)} \, {\Big( \frac{2 \pi X F^2}{T} \Big)}^X \, \exp \! \Big[-\frac{2 \pi X F^2}{T}
\Big] \, \Bigg\{ 1 - \frac{1}{2} \frac{T}{2 \pi F^2} + {\cal O} \Big( \frac{T^2}{F^4} \Big) \Bigg\} \, ,
\end{equation}
which depends on $N$ through the parameter $X$,
\begin{equation}
X = \frac{1}{N-2} \, .
\end{equation}
\begin{figure}
\includegraphics[width=13.5cm]{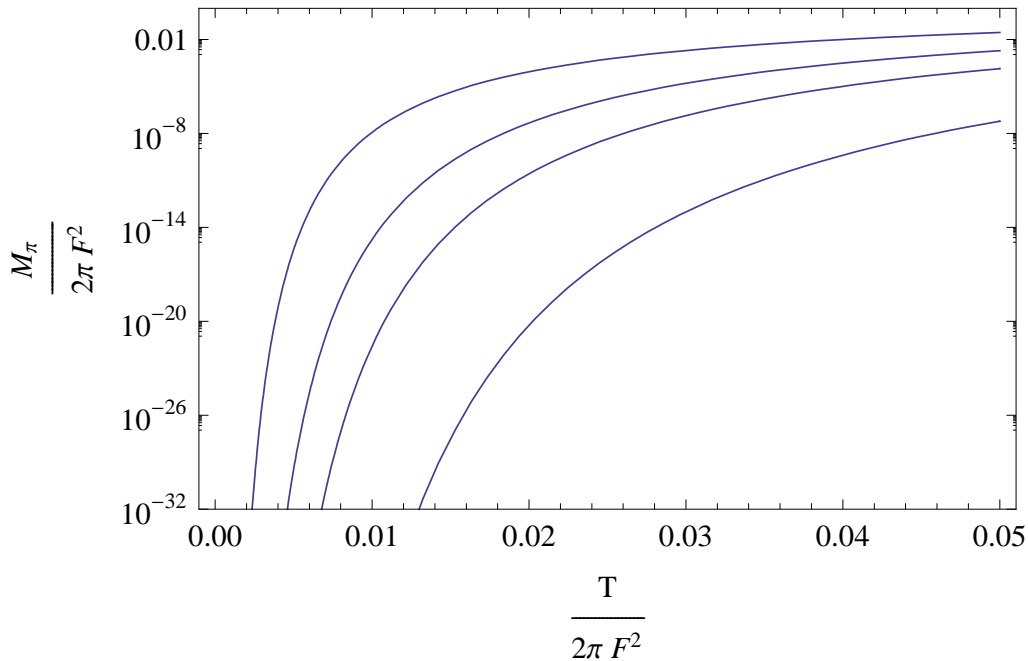}
\caption{Domain of validity of the effective low-temperature expansion of $d$=2+1 systems with a spontaneously broken symmetry O($N$) $\to$
O($N$-1). The effective expansion is safe in the parameter region above the curves that refer to $N$= \{3,4,5,8\}, from bottom to top in
the figure.}
\label{figure3}
\end{figure}
As discussed in detail in Ref.~\citep{Hof10}, the correlation length of the Goldstone bosons $\xi = 1/M_{\pi}$ must be much smaller than
the correlation length $\xi_{np}=1/m$ defined by the nonperturbatively generated mass, in order for the effective expansion to be valid.
This condition then leads to a relation between the two ratios $T/ 2 \pi F^2$ and $M_{\pi}/2 \pi F^2$, allowing one to identify the
parameter domain where the effective series are valid. This is illustrated for various values of $N$ in Fig.~\ref{figure3}. The region
below these curves corresponds to the parameter domain where the effective analysis presented here cannot be applied. While this excluded
region is tiny for $N$= \{3,4,5,8\}, it grows as $N$ increases.

We emphasize that the spontaneously broken internal symmetry O($N$) $\to$ O($N$-1) not necessarily refers to quantum spin systems. Just
for concreteness we continue using magnetic terminology, assuming that the underlying system is described by the $d$=2+1 antiferromagnetic
Heisenberg Hamiltonian for spins in arbitrary dimension $N$, and that the underlying scale can be identified with the exchange integral
$J$.

\subsection{Manifestation of the Interaction in the Pressure}
\label{InteractionPressure}

If the interaction contribution in the pressure has a positive (negative) sign, we are dealing with a repulsive (attractive) interaction.
In terms of the dimensionless ratio
\begin{equation}
\tau = \frac{T}{M_{\pi}} \, ,
\end{equation}
the pressure, up to three-loop order and for general $N$ ($N \ge 2$), reads
\begin{eqnarray}
\label{pressureTau}
P & = & \frac{N-1}{2} h_0(\sigma) T^3
- \frac{(N-1)(N-3)}{8 F^2 {\tau}^2} {h_1(\sigma)}^2 T^4 \nonumber \\
& + & \frac{(N-1)(N-3)(5N-11)}{128 \pi F^4 {\tau}^3} {h_1(\sigma)}^2 T^5
- \frac{(N-1)(N-3)(3N-7)}{48 F^4 {\tau}^2} {h_1(\sigma)}^3 T^5 \nonumber \\
& + & \frac{(N-1){(N-3)}^2}{16 F^4 {\tau}^4} {h_1(\sigma)}^2 h_2(\sigma) T^5
- \frac{1}{F^4} q(\sigma) T^5 + {\cal O}(T^6) .
\end{eqnarray}
As the system is homogeneous, the pressure is obtained from the free energy density, Eq.~(\ref{fedTau}), by
\begin{equation}
\label{Pz}
P = z_0 - z \, ,
\end{equation}
where $z_0$ is the free energy density at zero temperature.

\begin{figure}
\includegraphics[width=13.5cm]{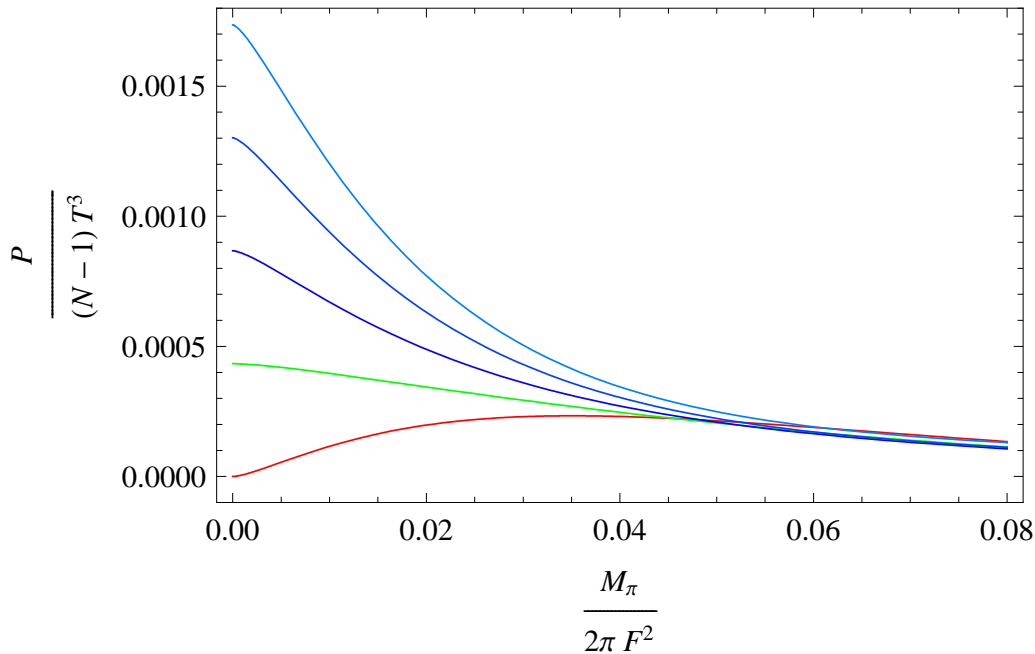}
\caption{[Color online] Manifestation of the interaction in the pressure of (2+1)-dimensional systems with a spontaneously broken symmetry
O($N$) $\to$ O($N$-1) for $N$= \{2,3,4,5,6\} from bottom to top in the figure (vertical cut at $M_{\pi}/2 \pi F^2 = 0.02$) in the figure.
The curves have been evaluated at $T/2 \pi F^2 = 0.05$.}
\label{figure4}
\end{figure}

The low-temperature series for the pressure allows us to extract the information on the Goldstone boson interaction we are interested in.
The following statements apply to (2+1)-dimensional (pseudo-)Lorentz-invariant systems with a spontaneously broken internal symmetry
O($N$) $\to$ O($N$-1). In Fig.~\ref{figure4} we have plotted the sum of the two-loop ($\propto T^4$) and three-loop ($\propto T^5$)
contributions to the pressure as a function of the ratio $M_{\pi}/2 \pi F^2$. In connection with spin systems, this explicit symmetry
breaking parameter measures the strength of the external staggered field. Note that we have scaled the pressure by the number of Goldstone
bosons that appear in a given system, i.e., we have plotted the dimensionless quantity
\begin{equation}
\frac{1}{N-1} \, \frac{P}{T^3} \, .
\end{equation}
For the temperature $T/2 \pi F^2 = 0.05$ that we have chosen in Fig.~\ref{figure4}, the interaction in the pressure is repulsive for any
of the systems $N$= \{2,3,4,5,6\} considered.

\begin{figure}
\includegraphics[width=13.5cm]{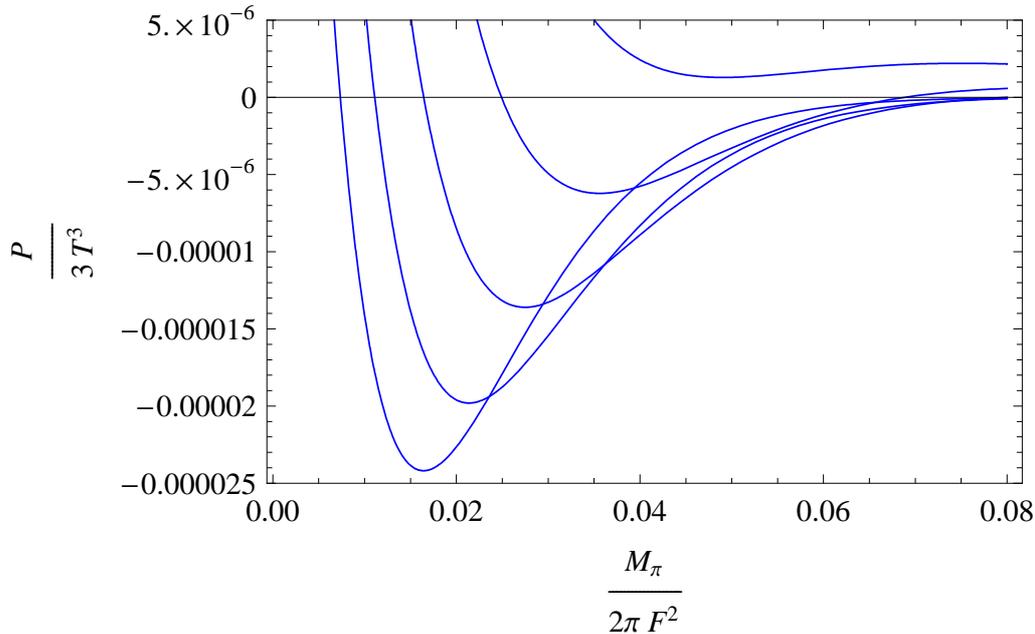}
\caption{Manifestation of the interaction in the pressure of (2+1)-dimensional systems with a spontaneously broken symmetry O(4) $\to$
O(3). The curves have been evaluated at the temperatures $T/2 \pi F^2 = \{ 0.028,\ 0.025,\ 0.022,\ 0.019,\ 0.016 \}$, from top to bottom
in the figure.}
\label{figure5}
\end{figure}

However, the situation becomes more subtle if we go to lower temperatures. For $N$=\{2,3\}, the interaction remains repulsive,
independently of the ratio $T/2 \pi F^2$. But the behavior of the systems with $N \ge 4$ is more complicated as we illustrate in
Fig.~\ref{figure5} for the case $N$=4. By reducing the temperature, the Goldstone boson interaction in the pressure becomes attractive in
an intermediate region that depends on the parameter $M_{\pi}/2 \pi F^2$. This region keeps growing as the temperature drops.

\begin{figure}
\includegraphics[width=13.5cm]{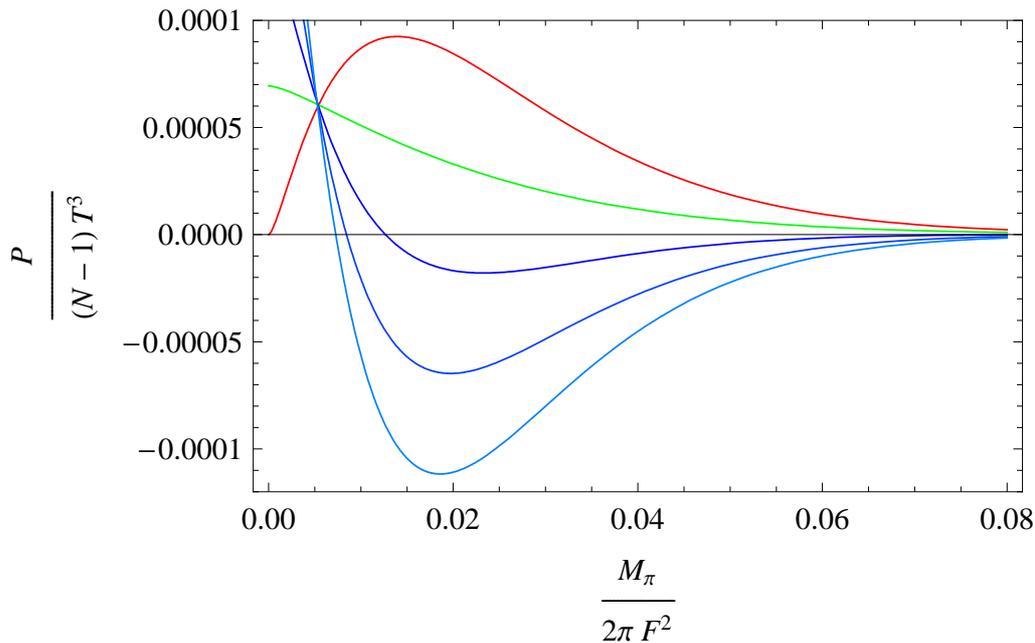}
\caption{[Color online] Manifestation of the interaction in the pressure of (2+1)-dimensional systems with a spontaneously broken symmetry
O($N$) $\to$ O($N$-1) for $N$= \{2,3,4,5,6\} from top to bottom in the figure (vertical cut at $M_{\pi}/2 \pi F^2 = 0.02$). The curves
have been evaluated at $T/2 \pi F^2 = 0.02$.}
\label{figure6}
\end{figure}

These effects are an immediate consequence of the $N$-dependence of the low-temperature expansion for the pressure. At the two-loop level,
there is just one term ($\propto T^4$) in Eq.~(\ref{pressureTau}). As it involves a factor of $N$-3, one may identify three classes of
systems regarding the Goldstone boson interaction: the Heisenberg antiferromagnet ($N$=3) where the interaction term vanishes, the
quantum XY model ($N$=2), where the interaction is repulsive, and systems with $N \ge 4$ where the interaction in the pressure is
attractive at two-loop order. Then, at the three-loop level, the $N$-dependence is more complicated, but overall leads to positive
contributions to the pressure. The Heisenberg antiferromagnet receives a repulsive contribution, while the repulsive interaction in
the quantum XY model gets enhanced. For systems with $N \ge 4$, the three-loop corrections are quite large at low temperatures, and
involve opposite signs compared to the two-loop contribution. This leads to the interesting behavior we discussed before, and which we
have depicted again in Fig.~\ref{figure6} for the temperature $T/2 \pi F^2 = 0.02$ for $N$= \{2,3,4,5,6\}. For quantum spin systems with
$N \ge 4$, depending on the strength of the external staggered field and the temperature, the Goldstone boson the interaction in the
pressure may be attractive, repulsive, or zero.

Note that the curves in Fig.~\ref{figure5} and Fig.~\ref{figure6} do not diverge in the limit $M_{\pi}/2 \pi F^2 \to 0$, i.e., when the
external field is switched off: they tend to a finite value as in Fig.~\ref{figure4} where the temperature is more elevated. This value is
positive, except for $N$=2, where the Goldstone boson interaction tends to zero and the system behaves as ideal gas. While the limit
$M_{\pi}/2 \pi F^2 \to 0$ is mathematically well-defined in the series for the pressure, one should keep in mind that, due to the
Mermin-Wagner theorem, or the nonperturbatively generated mass gap, one eventually leaves the domain of validity of the effective
expansion if $M_{\pi}$ decreases. However, for the temperature $T/2 \pi F^2 = 0.02$ chosen in Fig.~\ref{figure6}, the parameter
$M_{\pi}/2 \pi F^2$ must be really tiny in order for the effective series to break down. This can be easily verified by inspecting
Fig.~\ref{figure3}. Therefore, although it is conceptually inconsistent to take the limit $M_{\pi}/2 \pi F^2 \to 0$, numerically the
pressure is not affected.

We thus emphasize that the effects described above occur in a parameter region where our approach perfectly applies. The systematic
effective field theory analysis predicts that for systems with $N \ge 4$, depending on the ratios $T/2 \pi F^2$ and
$M_{\pi}/2 \pi F^2$, the Goldstone boson interaction in the pressure may be repulsive, attractive or zero. On the other hand, for
$N$= \{2,3\}, the interaction always is repulsive.

Finally we mention a subtlety related to the quantities $M$ and $M_{\pi}$ that both depend on the external field $H_s$, and are related by
Eq.~(\ref{MassRen}). The point is that the interaction already manifests itself at zero temperature: in the presence of an external field,
the bare Goldstone boson mass $M = \sqrt{\Sigma_s H_s}/F$ receives one-loop and two-loop contributions: $M \to M_{\pi}$. In the figures
shown in the present subsection, we have plotted the {\it finite-temperature} interaction contribution in the pressure as a function of
the quantity $M_{\pi}/2 \pi F^2$ that involves $M_{\pi}$, and thus takes into account the effect of the interaction at zero temperature. The
figures for the pressure, that all refer to finite temperature, have thus to be interpreted as follows: we start at $T$=0 and switch on
the external field. Then we switch on the temperature (e.g., in Fig.~\ref{figure6} the temperature is $T/2 \pi F^2 = 0.02$), and observe
whether the interaction at {\it finite temperature} is repulsive, attractive, or zero.

Note again that the systems under consideration not necessarily are quantum spin models -- the effects discussed above apply to any
(pseudo-)Lorentz-invariant system in two spatial dimensions exhibiting a spontaneously broken internal symmetry O($N$) $\to$ O($N$-1). We
find it remarkable that the effects concerning the nature of the interaction follow from symmetry considerations alone: they neither
depend on the concrete realization of the system nor on its microscopic details.

\subsection{Manifestation of the Interaction in the Staggered Magnetization}
\label{InteractionOrderParameter}

The staggered magnetization at finite temperature is defined by
\begin{equation}
\Sigma_s(T,H_s) = - \frac{\partial z}{\partial H_s} .
\end{equation}
It represents the order parameter, signaling that the internal rotation symmetry is spontaneously broken.\footnote{Strictly speaking,
there is no spontaneous symmetry breaking at finite temperature and zero external field in two spatial dimensions due to the theorem by
Mermin-Wagner. But the staggered magnetization still takes nonzero values at finite and low temperatures and weak external field. In this
sense we refer to the quantity $\Sigma_s(T,H_s)$ as order parameter.} 

Using the expression for the free energy density (\ref{fedTau}), the low-temperature expansion takes the form 
\begin{eqnarray}
\label{OPTauN2}
\Sigma_s(T,H_s) & = & \Sigma_s(0,H_s) - \frac{(N-1)}{2} \frac{\Sigma_s b}{F^2} h_1(\sigma) T \nonumber \\
& - & \frac{(N-1)(N-3)}{8} \frac{\Sigma_s b}{F^4} \Big\{ {h_1(\sigma)}^2 - \frac{2}{\tau^2} h_1(\sigma) h_2(\sigma) \Big\} T^2 \nonumber \\
& + & \frac{(N-1)(N-3)(5N-11)}{128 \pi} \frac{\Sigma_s b}{F^6} \Big\{ {\frac{3}{2 \tau} h_1(\sigma)^2}
- \frac{2}{\tau^3} h_1(\sigma) h_2(\sigma) \Big\} T^3 \nonumber \\
& - & \frac{(N-1)(N-3)(3N-7)}{48} \frac{\Sigma_s b}{F^6} \Big\{ {h_1(\sigma)}^3 -\frac{3}{\tau^2} {h_1(\sigma)}^2 h_2(\sigma) \Big\} T^3
\nonumber \\
& + & \frac{(N-1){(N-3)}^2}{16} \frac{\Sigma_s b}{F^6} \Big\{ \frac{2}{\tau^2} {h_1(\sigma)}^2 h_2(\sigma) -
\frac{2}{\tau^4} h_1(\sigma) {h_2(\sigma)^2} - \frac{1}{\tau^4} {h_1(\sigma)}^2 h_3(\sigma) \Big\} T^3 \nonumber \\
& - & \frac{\Sigma_s b}{8 \pi^2 F^6 \sigma} \frac{\partial q(\sigma)}{\partial \sigma } T^3 + {\cal O}(T^4) ,
\end{eqnarray}
where $b$ is given by
\begin{eqnarray}
\label{defb}
b(H_s) & = & \frac{\partial M^2_{\pi}}{\partial M^2} = 1 - \frac{3 (N-3) \sqrt{\Sigma_s}}{16 \pi F^3} \sqrt{H_s}
+ \frac{2 \tilde k_0 \Sigma_s}{F^4} H_s + {\cal O}(H_s^{3/2}) , \nonumber \\
& & \tilde k_0 = 2 (k_2 - k_1) + \frac{b_1}{F^2} + \frac{b_2}{64 \pi^2 F^2} .
\end{eqnarray}
The next-to-leading order effective constants $k_1$ and $k_2$ are defined in Eq.~(\ref{Leff}), while the explicit expressions for the
coefficients $b_1$ and $b_2$ can be found in Appendix B of Ref.~\citep{Hof10}.

\begin{figure}
\includegraphics[width=12.0cm]{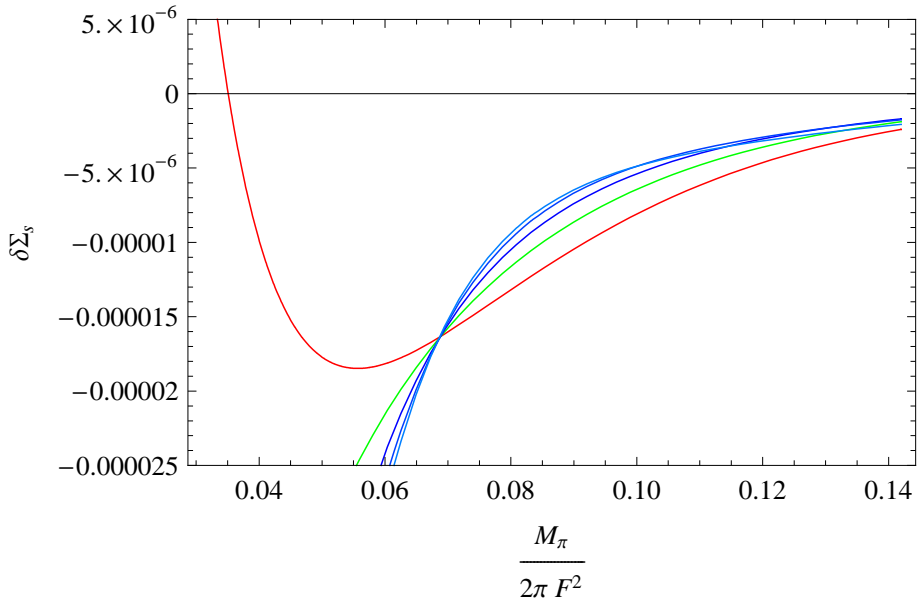}
\caption{[Color online] Manifestation of the temperature-dependent interaction contribution in the order parameter (staggered
magnetization) of (2+1)-dimensional systems with a spontaneously broken symmetry O($N$) $\to$ O($N$-1) for $N$= \{2,3,4,5,6\} from bottom
to top in the figure (vertical cut at $M_{\pi}/2 \pi F^2 = 0.08$). The curves have been evaluated at $T/2 \pi F^2 = 0.05$.}
\label{figure9}
\end{figure}

In the two figures of this subsection, we plot the sum of the two- and three-loop contributions in the dimensionless quantity
\begin{equation}
\delta \Sigma_s \equiv \frac{1}{N-1} \frac{\Sigma_s(T,H_s) - \Sigma_s(0,H_s)}{\Sigma_s} ,
\end{equation}
which measures the strength and the sign of the Goldstone boson interaction in the temperature-dependent part of the staggered
magnetization. Note that the $T$-{\it independent} part of the staggered magnetization in the presence of an external field is given by
\begin{eqnarray}
\label{SigmaN}
\Sigma_s(0,H_s) & = & \Sigma_s + \frac{(N-1) \, \Sigma_s^{3/2}}{8 \pi F^3} \sqrt{H_s} + {\cal O}(H_s) , \nonumber \\
\Sigma_s & = & \Sigma_s(0,0) .
\end{eqnarray}
Since we are at zero temperature, the limit $H_s \to 0$ in Eq.~(\ref{SigmaN}) raises no problems: the Mermin-Wagner theorem does not
apply.

As can be appreciated in Fig.~\ref{figure9}, the $d$=2+1 quantum XY model (or any other physical realization of $N$=2) is quite different
from systems with $N \ge 3$: while the sign of the interaction is negative in most of the parameter domain, for small external fields
(small $M_{\pi}$) it is positive. For all other systems ($N \ge 3$) the temperature-dependent interaction contribution in the order
parameter tends to large negative values if the external field is weak.

\begin{figure}
\includegraphics[width=12.0cm]{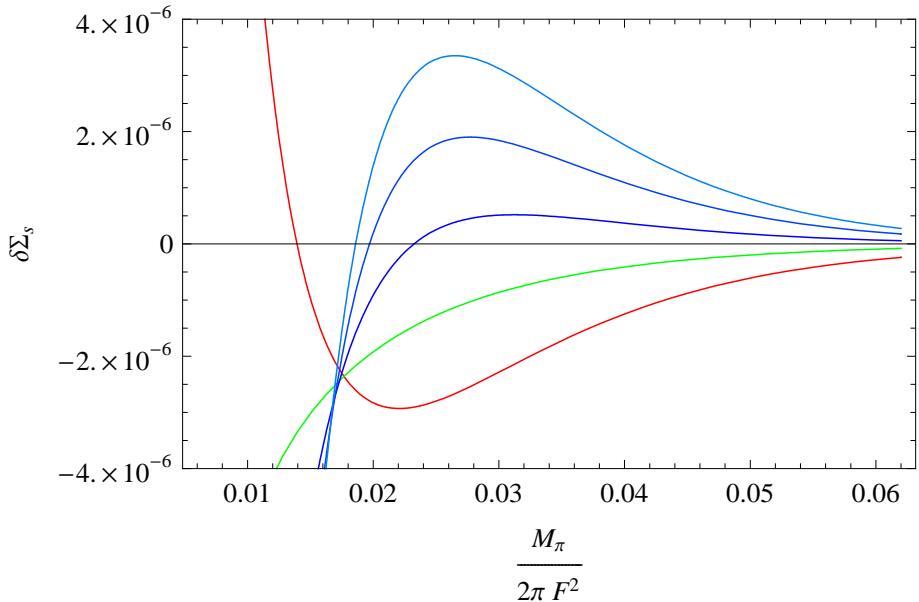}
\caption{[Color online] Manifestation of the temperature-dependent interaction contribution in the order parameter (staggered
magnetization) of (2+1)-dimensional systems with a spontaneously broken symmetry O($N$) $\to$ O($N$-1) for $N$= \{2,3,4,5,6\} from bottom
to top in the figure (vertical cut at $M_{\pi}/2 \pi F^2 = 0.03$). The curves have been evaluated at $T/2 \pi F^2 = 0.02$.}
\label{figure9b}
\end{figure}

Note that Fig.~\ref{figure9} corresponds to $T/2 \pi F^2 = 0.05$. If we go to lower temperatures, effects similar to those observed for
the pressure take place: while the properties of the quantum XY model and the Heisenberg antiferromagnet qualitatively remain the same,
systems with $N \ge 4$ develop positive values for the interaction part in the staggered magnetization, as we illustrate in
Fig.~\ref{figure9b} that refers to $T/2 \pi F^2 = 0.02$. In particular, since the curves for $N \ge 4$ now exhibit a maximum, the
staggered susceptibility for these systems vanishes at this point in parameter space (see next subsection). Remarkably, in the case of the
Heisenberg antiferromagnet, the quantity $\delta \Sigma_s$ always takes negative values and never presents an extremum, independently of
the temperature or the external field.

The subleading effects discussed here are quite small, and the behavior of the staggered magnetization is in fact dominated by the
one-loop contribution. As one would expect, the one-loop contribution in $\delta \Sigma_s$ is negative for any of the systems: the
corresponding term, proportional to one power of the temperature in Eq.~(\ref{OPTauN2}), is governed by $- h_1$ (which is negative), where
$h_1$ is the kinematical function that we have depicted in Fig.~\ref{kinematical}. If the temperature is switched on while the external
field is held fixed ($M_{\pi}$ constant), indeed the order parameter decreases. Remember that we first switch on the external field $H_s$
at $T$=0, and then go to finite temperature, keeping $H_s$ fixed.

Interestingly, the effect of the Goldstone boson interaction, as witnessed by Fig.~\ref{figure9} and Fig.~\ref{figure9b}, does not
necessarily follow this behavior observed for the one-loop contribution. For the $d$=2+1 quantum XY model at low temperatures and weak
external field, the effect is just the opposite: the interaction at finite temperature leads to an increase of the staggered
magnetization, meaning that here the interaction tends to reinforce the antialignment of the spins and enhances the staggered spin
pattern. Similarly, for systems with $N \ge 4$, the interaction leads to a slight increase of the order parameter at low temperatures for
not too weak external fields (see Fig.~\ref{figure9b}).  Overall, of course, the one-loop contribution dominates, such that the order
parameter still decreases when the temperature is switched on.

It should be pointed out that the low-temperature series for the staggered magnetization -- in contrast to the series for the pressure --
formally diverges if the external field is switched off. As we have explained before, taking this limit in any observable is problematic,
as we leave the domain of validity of the effective expansion. Taking this limit in the pressure did not numerically affect the series.
However, in the staggered magnetization and the staggered susceptibility (see below), this limit is divergent. But note that the curves
displayed in Fig.~\ref{figure9} and Fig.~\ref{figure9b} can be trusted to very low values of $M_{\pi}/2 \pi F^2$ (very weak external
field) according to Fig.~\ref{figure3}, such that the finite-temperature interaction contribution in the order parameter indeed tends to
large positive (negative) values for $N$=2 ($N \ge 3$) in a regime where the effective expansion still applies.

At the end of this section, let us explain how we have generated the curves concerning the range of validity of the effective expansion
that we have depicted in Fig.~\ref{figure2}. For a fixed value of the external field $H_s$, the order parameter $\Sigma_s(T,H_s)$ drops
to zero when the temperature is raised sufficiently: at this point the effective expansion certainly no longer applies -- we are now in a
region of elevated temperature where the Goldstone boson picture ceases to be valid. To estimate this "critical" temperature for a given
constant value of $H_s$, we take the first two terms in the expansion for the staggered magnetization, Eq.~(\ref{OPTauN2}), and solve the
equation
\begin{equation}
\label{checkField}
\Sigma_s(T,H_s) = \Sigma_s + \frac{N-1}{8 \pi} \, \frac{\Sigma_s^{3/2}}{F^3} \sqrt{H_s}
- \frac{N-1}{2} \, \frac{\Sigma_s b}{F^2} h_1(\sigma) \, T \equiv 0 \, ,
\end{equation}
with the kinematical function $h_1$
\begin{equation}
h_1(\sigma) = - \frac{1}{2 \pi} \ln \Big( 1-e^{-M_{\pi}/T} \Big) \approx - \frac{1}{2 \pi} \ln \Big( 1-e^{-\sqrt{\Sigma_s H_s}/FT} \Big) \, .
\end{equation}
For a given ratio of $H_s/2 \pi F^2$ (i.e., $H_s/J$) we thus obtain a "critical" temperature $T_c/J$ that signals the breakdown of the
effective expansion. While in Fig.~\ref{figure2} we have shown the situation for the $d$=2+1 quantum XY model and the Heisenberg
antiferromagnet (where the specific values for $F^2$, $\Sigma_s$ and the spin-wave velocity $v_s$ have been measured), the situation for
$N \ge 4$ is similar: the effective expansion only breaks down when the external field is tiny, as depicted in Fig.~\ref{figure3}.

\subsection{Manifestation of the Interaction in the Staggered Susceptibility}
\label{InteractionSusceptibility}

In connection with spin systems, the staggered susceptibility 
\begin{equation}
\chi(T,H_s) = \frac{\partial \Sigma_s(T,H_s)}{\partial H_s} = \frac{\Sigma_s}{F^2} \, \frac{\partial \Sigma_s(T,M)}{\partial M^2}
\end{equation}
is the response of the staggered magnetization to the applied external field. Rather than providing lengthy expressions for $\chi(T,H_s)$
that can trivially be obtained from the low-temperature expansion of the order parameter, Eq.~(\ref{OPTauN2}), we just exhibit the the
general structure of the series:
\begin{equation}
\label{seriesSuscept}
\chi(T,H_s) = \chi(0,H_s) + \chi_1(\tau) \frac{1}{T} + \chi_2(\tau) + \chi_3(\tau) T + {\cal O}(T^2) \, .
\end{equation}
The coefficients $\chi_i$, much like the analogous coefficients in the pressure and the staggered magnetization, depend in a nontrivial
way on the dimensionless ratio $\tau = T/M_{\pi}$.

Regarding quantum spin systems, the staggered field ${\vec H_s}=(0,\dots, 0,H_s)$ tends to align the spins in an antiparallel pattern and
hence increases the staggered magnetization. One thus expects the staggered susceptibility $\chi(T,H_s)$ in general to be positive. This
is indeed true for the noninteracting (one-loop) contribution in Eq.~(\ref{seriesSuscept}), i.e., the leading temperature-dependent term
that involves an inverse power of $T$. However, the effects related to the interaction are more subtle.

\begin{figure}[t]
\includegraphics[width=13.5cm]{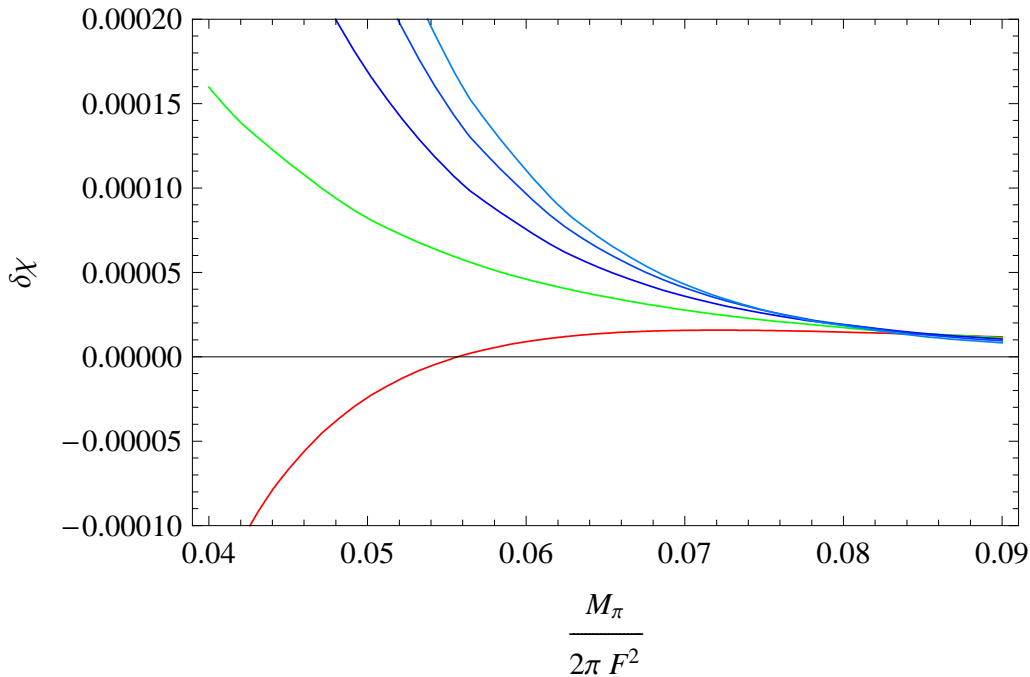}
\caption{[Color online] Manifestation of the temperature-dependent interaction contribution in the staggered susceptibility of
(2+1)-dimensional systems with a spontaneously broken symmetry O($N$) $\to$ O($N$-1) for $N$= \{2,3,4,5,6\} from bottom to top in the
figure (vertical cut at $M_{\pi}/2 \pi F^2 = 0.06$). The curves have been evaluated at $T/2 \pi F^2 = 0.05$.}
\label{figure10}
\end{figure}

\begin{figure}[t]
\includegraphics[width=13.5cm]{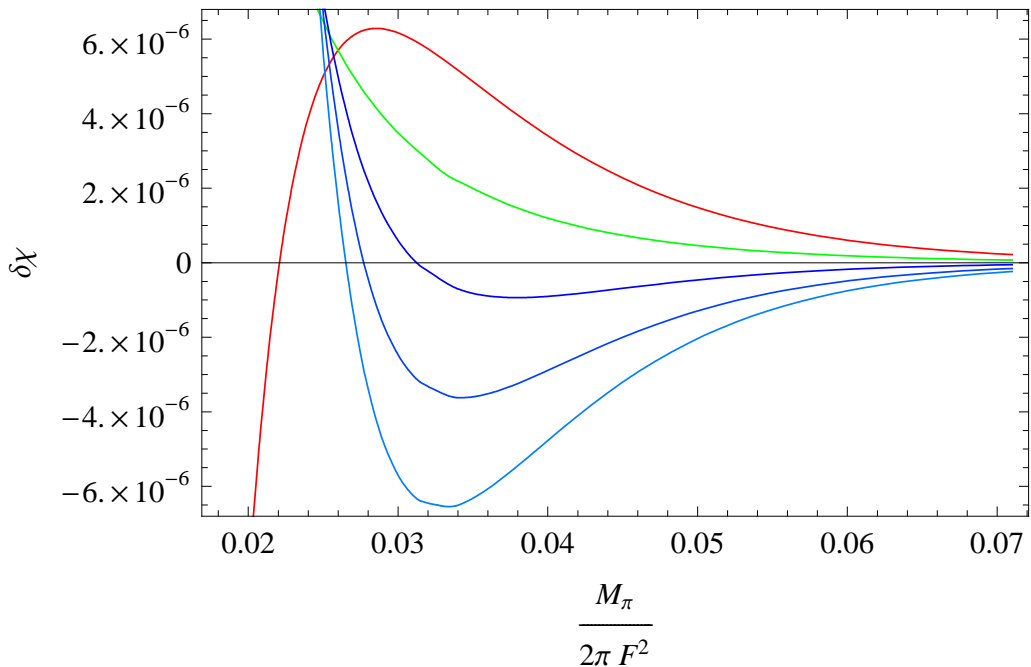}
\caption{[Color online] Manifestation of the temperature-dependent interaction contribution in the staggered susceptibility of
(2+1)-dimensional systems with a spontaneously broken symmetry O($N$) $\to$ O($N$-1) for $N$= \{2,3,4,5,6\} from top to bottom in the
figure (vertical cut at $M_{\pi}/2 \pi F^2 = 0.04$). The curves have been evaluated at $T/2 \pi F^2 = 0.02$.}
\label{figure11}
\end{figure}

In the subsequent two figures, we plot the sum of the two- and three-loop contributions in the dimensionless quantity
\begin{equation}
\delta \chi \equiv \frac{T F^4}{N-1} \, \frac{\chi(T,H_s) - \chi(0,H_s)}{\Sigma^2_s} ,
\end{equation}
which measures the strength and the sign of the Goldstone boson interaction in the temperature-dependent part of the staggered
susceptibility.

The behavior of the $d$=2+1 quantum XY model is peculiar, as the finite-temperature interaction contribution in the staggered
susceptibility -- in contrast to all other systems ($N \ge 3$) -- becomes negative for small values of $M_{\pi}/2 \pi F^2$, as shown in
Fig.~\ref{figure10} and Fig.~\ref{figure11}. The $d$=2+1 Heisenberg antiferromagnet also plays a special role as it is the only system
with a positive interaction contribution in the staggered susceptibility, independent of the strength of the staggered field (or magnitude
of $M_{\pi}/2 \pi F^2$) and temperature. For $N \ge 4$ the behavior at $T/2 \pi F^2 = 0.05$ is qualitatively the same as for the
antiferromagnet (see Fig.~\ref{figure10}), but for lower temperatures the quantity $\delta \chi$ becomes negative for not too weak
external fields (see Fig.~\ref{figure11}). But note that for small values of $M_{\pi}/2 \pi F^2$, the staggered susceptibility ($N \ge 3$)
tends to large positive values for any temperature -- the $d$=2+1 quantum XY model really is the exception.

Analogous to the staggered magnetization, the manifestation of the Goldstone boson interaction in the staggered susceptibility is thus
quite subtle. In many cases the temperature-dependent interaction contribution in $\chi(T,H_s)$ is positive, implying that the external
field reinforces the staggered spin pattern. In some cases, however, the $T$-dependent interaction part in $\chi(T,H_s)$ is negative,
meaning that here the effect of the staggered field is to perturb the antialignment of the spins and to decrease the staggered
magnetization -- or the order parameter in general. This happens for the $d$=2+1 quantum XY model at weak external field (and finite
temperature), and for systems with $N \ge 4$ at low temperatures and not too weak external fields. It should be pointed out again that
these subtle effects regarding the nature of the interaction at finite temperature are rather weak -- the noninteracting part in fact
dominates, and leads to positive staggered susceptibilities, as one would expect.

\section{Conclusions}
\label{Summary}

Our main results concern the universal nature of the Goldstone boson interaction in $d$=2+1 (pseudo-)Lorentz-invariant systems with a
spontaneously broken internal rotation symmetry O($N$) $\to$ O($N$-1). Our effective field theory calculations are valid at low
temperatures and in presence of a weak external field that explicitly breaks internal rotation invariance and thus invalidates the
Mermin-Wagner theorem. As concrete physical realizations we have considered quantum spin systems.

The $d$=2+1 quantum XY model and Heisenberg antiferromagnet -- or any other physical realizations of $N = \{ 2,3 \}$ -- are rather
special, since the interaction in the pressure is always repulsive, independent of the ratio $T/2 \pi F^2$ and $M_{\pi}/2 \pi F^2$ that
measure the temperature and the strength of the external field, respectively. For all other systems ($N \ge 4$), the interaction in the
pressure may be repulsive, attractive, or zero, depending on the strength of the external field and the temperature that hence control the
nature of the Goldstone boson interaction.

Regarding the staggered magnetization $\Sigma_s(T,H_s)$ -- the order parameter -- the interaction may lead to interesting and subtle
effects. When the temperature is switched on while keeping the strength of the external field fixed, in most scenarios the interaction
tends to decrease the staggered magnetization $\Sigma_s(T,H_s)$, as one would expect. However, the $d$=2+1 quantum XY model represents an
exception, because in very weak external fields and at finite temperature, the interaction tends to reinforce the staggered pattern,
rather than to destroy the antialignment of the spins. Systems with $N \ge 4$ exhibit a similar behavior at very low temperatures in case
the external field is not too weak: the order parameter is increased in this parameter regime, when we go from zero to finite temperature
while keeping the staggered field constant.

These effects also manifest themselves in the staggered susceptibility $\chi(T,H_s)$. In weak external fields, the temperature-dependent
interaction part in $\chi(T,H_s)$ tends to positive values for $N \ge 3$, but in the case of the $d$=2+1 quantum XY model it becomes
negative. For systems with $N \ge 4$, the $T$-dependent interaction contribution in the staggered susceptibility takes positive values,
but for very low temperatures it may drop to negative values provided that the external field is not too weak.

Although the calculation was performed within a (pseudo-)Lorentz-invariant framework, it readily applies to the $d$=2+1 quantum XY model
or the $d$=2+1 Heisenberg antiferromagnet. The crucial point is that the effective theory predictions regarding the nature of the
Goldstone boson interaction are not affected by Lorentz-non-invariant terms, that would indeed be present in the next-to-leading order
Lagrangian ${\cal L}^4_{eff}$, and that have not been taken into account. These terms are only relevant for a one-loop graph and are
contained in the temperature-independent quantity $M_{\pi}$. Only in this trivial way, and only at next-to-next-to-leading order, do these
anisotropies emerge in our expressions. In particular, they cannot change the sign of the interaction -- and hence alter our conclusions
-- in any of the observables considered. In this sense, the effective field theory predictions are universal or model-independent. We find
it indeed quite remarkable that the nature of the Goldstone boson interaction is fully determined by the symmetries of the system.

While we have explicitly referred to the quantum XY model and the Heisenberg antiferromagnet -- or spin systems characterized by higher
values $N$ of spin vector components -- the results presented here go beyond quantum spin models. Our low-temperature series are valid for
any $d$=2+1 (pseudo-)Lorentz-invariant system with a spontaneously broken internal rotation symmetry O($N$) $\to$ O($N$-1) in presence of a
weak external field. It is our hope that these subtle effects may be confirmed in future studies, e.g., by Monte Carlo simulations.

\section*{Acknowledgments}

The author would like to thank H.\ Leutwyler and U.-J.\ Wiese for stimulating discussions.

\end{document}